\documentstyle[psfig,conf_iap,]{article}
\begin{document}
\heading{%
%
Parsec-scale structure in galactic disk and halo, from diffuse radio
polarization
%
} 
\par\medskip\noindent
\author{%
Marijke Haverkorn$^{1}$, Peter Katgert$^{1}$, Ger de Bruyn$^2$
} 

\address{Leiden Observatory, P.O. Box 9513, 2300 RA Leiden, the
Netherlands\\
$^2$ASTRON, P.O. Box 2, 7990 AA Dwingeloo, the Netherlands}

\begin{abstract}
  We present multi-frequency polarization observations of the diffuse
  radio synchrotron background modulated by Faraday rotation. No total
  intensity is observed, indicating that total intensity does not vary
  on scales below approximately a degree. However, polarized intensity
  and polarization angle show abundant small-scale structure due to
  Faraday rotation in the warm ionized disk on small scales. The
  distribution of Rotation Measures enables us to estimate structure
  in magnetic field, weighted with electron density, in the warm
  ionized disk.
\end{abstract}

\section{The observations}

Synchrotron radiation emitted in our galaxy provides a diffuse radio
background. Small-scale structure in the linearly polarized component
of this radio background provides unique information about the
structure in the warm Interstellar Medium (ISM) and the galactic halo
on parsec scales and larger (see discovery paper \cite{wieringa}).

With the Westerbork Synthesis Radio Telescope (WSRT) we mapped the
polarized radio background in a 7$^{\circ}\times 9^{\circ}$ field
centered on $(l,b) = (161, 16)^{\circ}$, in five frequency bands (341,
349, 355, 360, and 375~MHz) with a bandwidth of 5 MHz simultaneously,
at a resolution of $\sim$4$^{\prime}$.  No total intensity $I$ was
detected down to $\sim$0.7~K, which is $\lsim 1.5\%$ of the expected
sky brightness in this region, indicating that $I$ does not vary on
scales detectable to the interferometer, i.e.\ below about a degree.
However, linearly polarized intensity $P$ and polarization angle
$\phi$ show abundant small-scale structure.

\subsection{Structure in polarized intensity $P$}
\begin{figure}
\centerline{\vbox{\psfig{figure=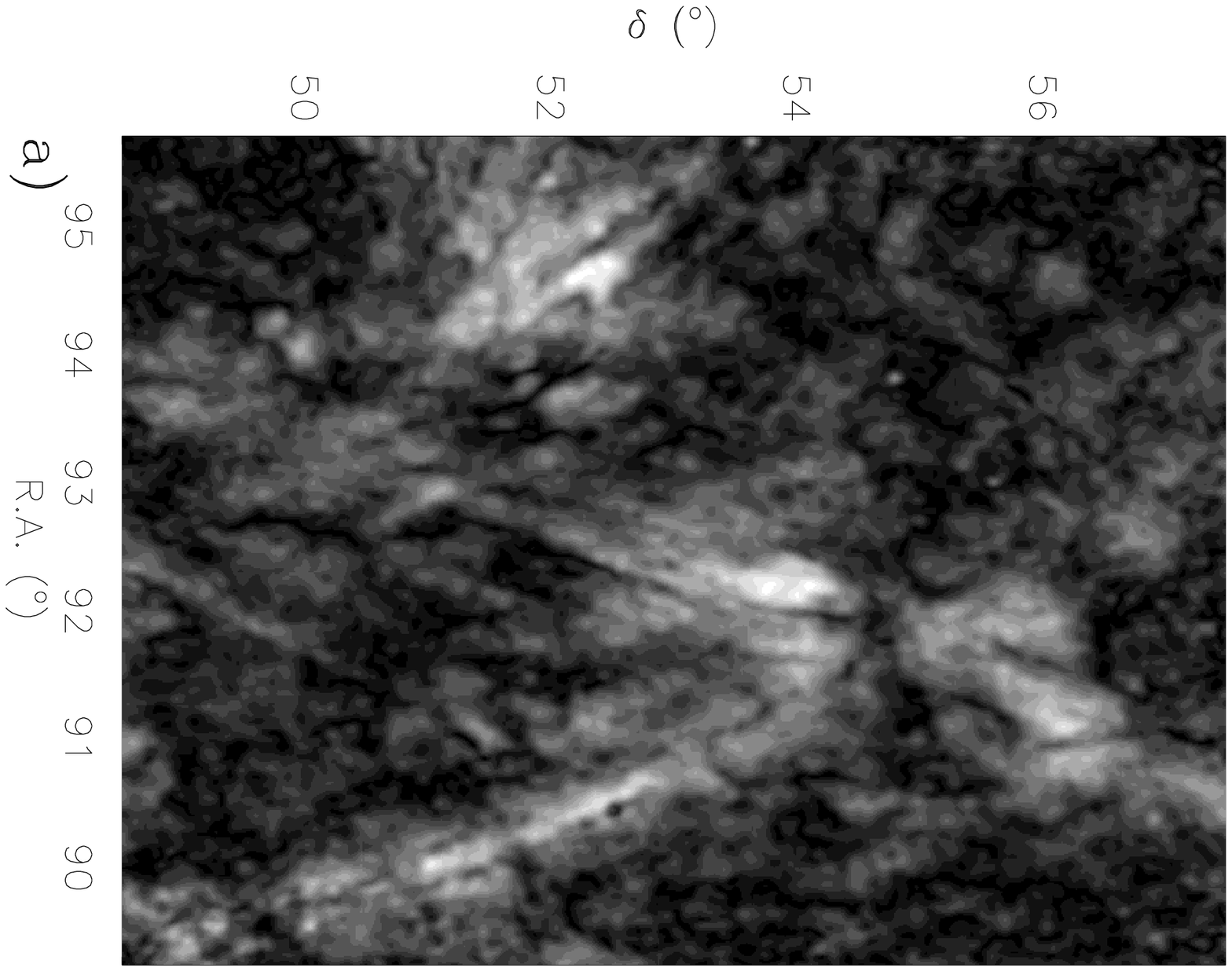,angle=90,width=0.47\textwidth}}
             \psfig{figure=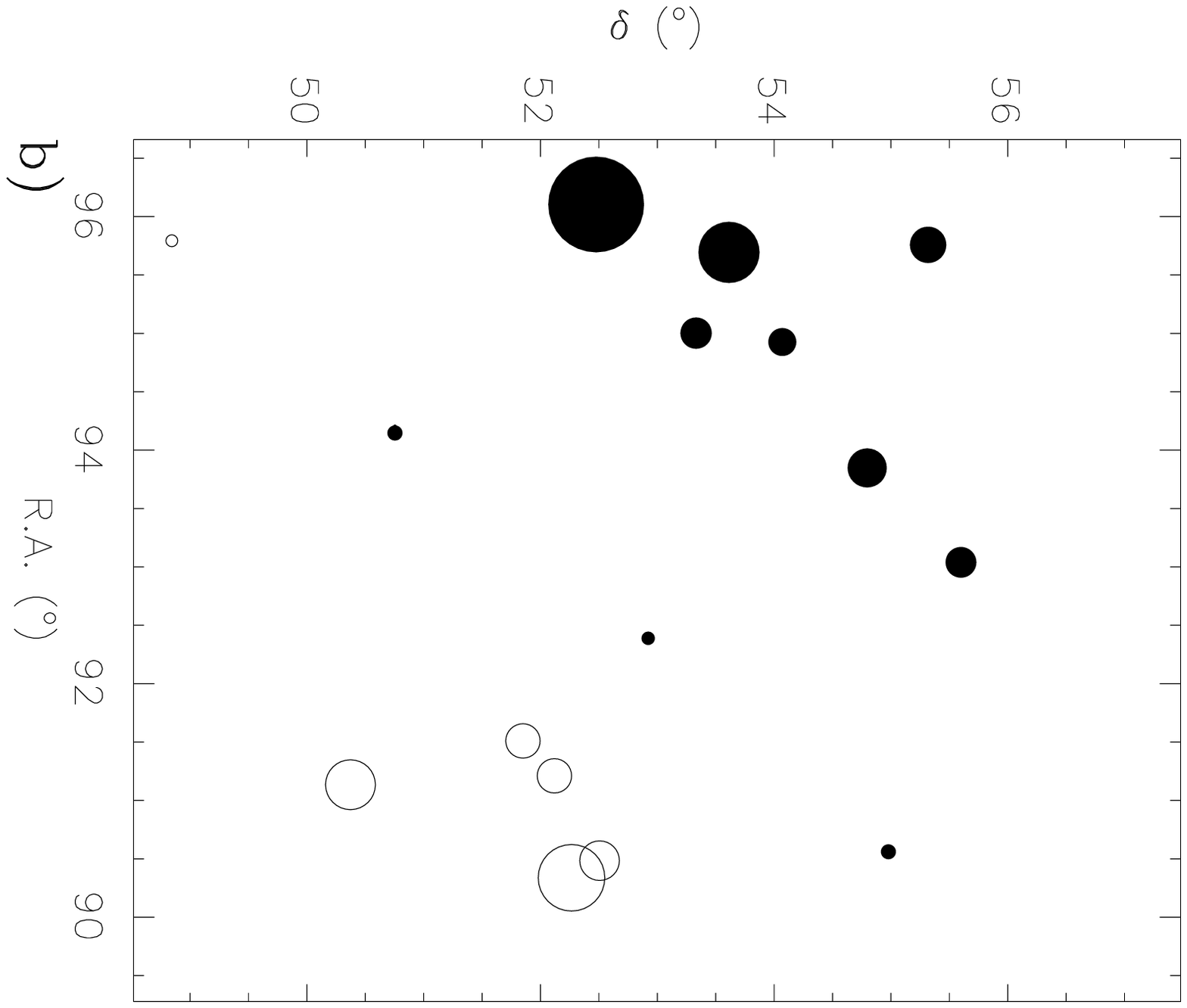,angle=90,width=0.50\textwidth}}
\caption[]{a) Polarized intensity map at 349~MHz at 4$^{\prime}$
resolution, white denotes a maximum $T_{b,pol}\approx 15$ K; b)
Rotation Measures (RM's) of observed polarized extragalactic
sources. The radii of the circles are scaled with magnitude of RM,
where black circles denote positive RM's. Maximum (minimum) RM is 19.5
($-13.6$) rad m$^{-2}$.}
\label{fig1}
\end{figure}

There is a wide variety in topology of the structure in polarized
intensity $P$, on several scales (Fig.~\ref{fig1}a).  The typical
brightness temperature $T_{b,pol} \approx$ 6 - 8 K (maximum
$\sim$15~K), and the maximum degree of polarization is $\sim$30\%,
with an average of $\sim$10\%. In addition, a pattern of black narrow
wiggly canals is visible (see e.g.\ the canal around
(RA,~dec) = (92.7, 49 - 51)$^{\circ}$), caused by beam depolarization.
These canals are all exactly one synthesized beam wide and have been
shown to be borders between regions of fairly constant polarization
angle $\phi$ where the difference in $\phi$ is approximately
90$^{\circ}$ ($\pm\, n\, 180^{\circ}$, $n=1,2,3\ldots$) caused by
abrupt changes in Rotation Measure (RM) \cite{haverkorn}.  Hence, the
canals reflect specific features in the angle distribution.  Other
angle (and RM) changes within the beam cause less or no
depolarization, so that they do not leave easily visible traces in the
polarized intensity distribution.

\subsection{Extragalactic sources}

Seventeen polarized extragalactic sources were detected at a higher
resolution ($\sim$1$^{\prime}$), with RM's from $-13.6$ to 19.5
rad~m$^{-2}$.  Fig.~\ref{fig1}b shows the RM's and positions of the
sources, where the radii of the circles are proportional to RM, and
white (black) circles denote negative (positive) RM's. The RM's of the
extragalactic sources increase roughly in the direction of galactic
latitude, indicating a galactic component to the RM's of the
sources. We estimate a RM component intrinsic to the source of
$\lsim$5 rad m$^{-2}$, consistent with earlier estimates \cite{leahy}.
The RM of the diffuse galactic radiation at the position of an
extragalactic source is independent from the observed RM of the source
itself (see Sect.~\ref{sec_rm}).

\subsection{Structure in Rotation Measure}
\label{sec_rm}

\begin{figure}
\centerline{\psfig{figure=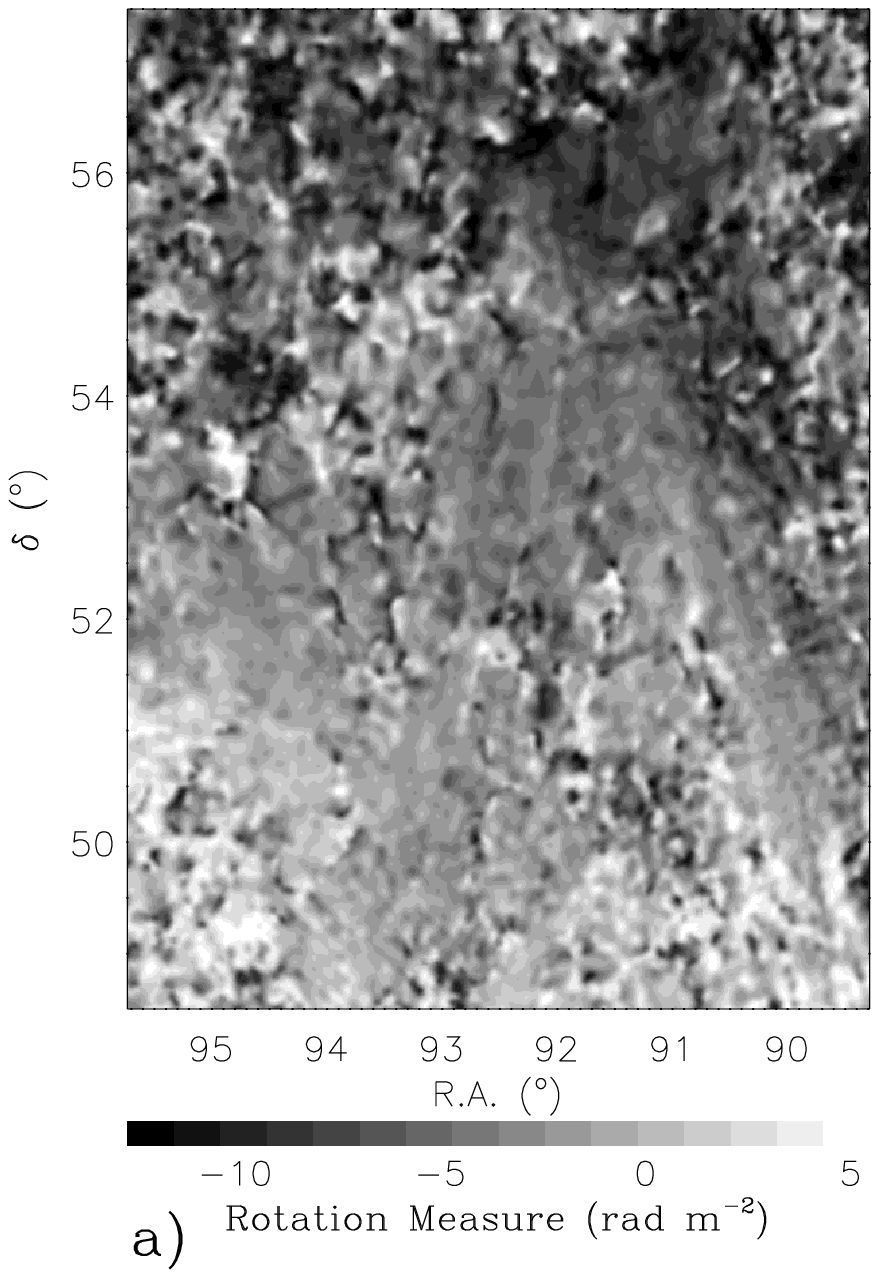,width=0.45\textwidth}
       \psfig{figure=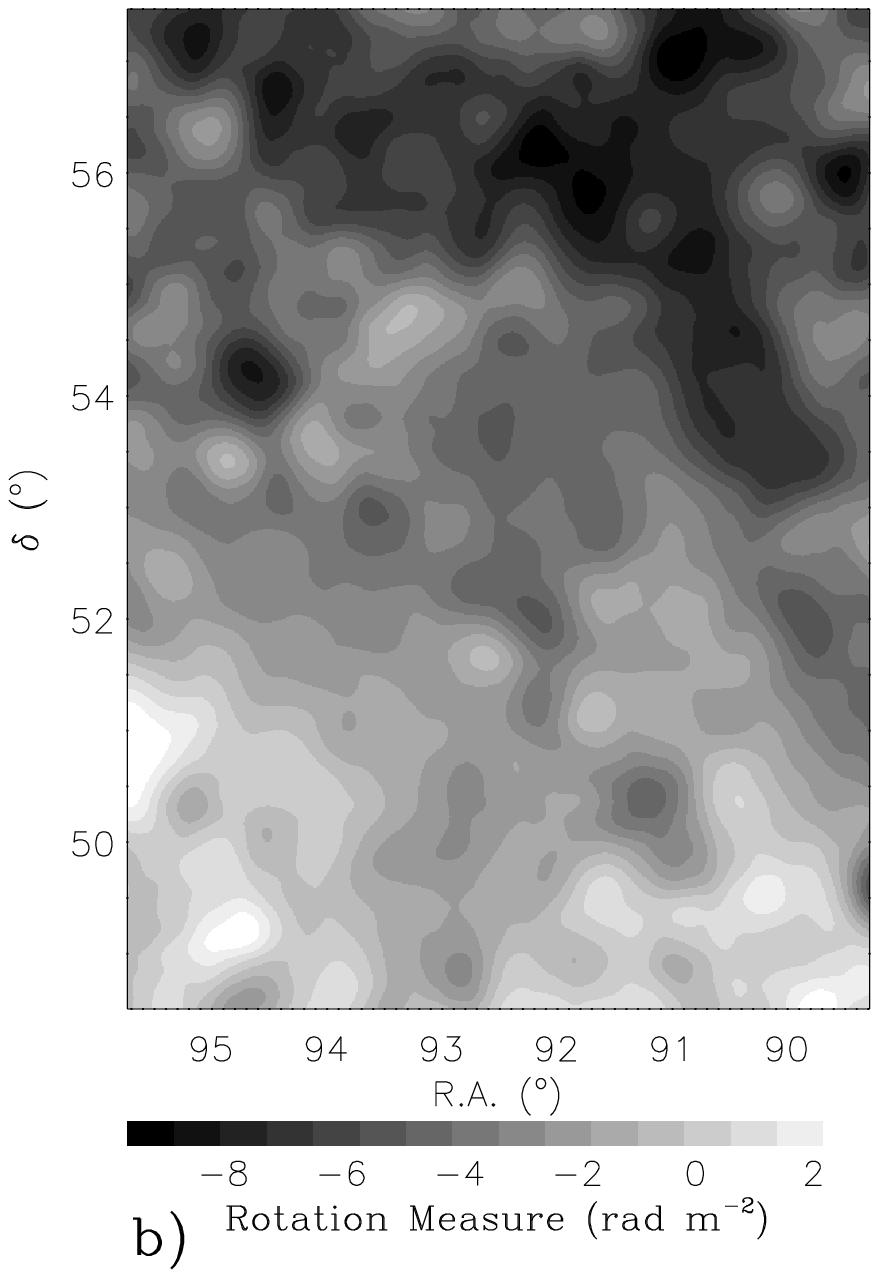,width=0.45\textwidth}}
\caption[]{RM maps: a) 5 times oversampled; b) smoothed over 8
  beams}
\label{fig2}
\end{figure}

The Rotation Measure of the Faraday-rotating material can be derived
from $\phi(\lambda^2) \propto \mbox{RM}\,\lambda^2$ (see
\cite{haverkorn2} for details and pitfalls).  Fig.~\ref{fig2}a gives
a 4$^{\prime}$ resolution RM map (which is oversampled), whereas in
Fig.~\ref{fig2}b, RM's are smoothed over 8 beams ($\sim
30^{\prime}$). The average RM $\approx -3.4$ rad~m$^{-2}$, and in
general $|\mbox{RM}| \lsim 10$ rad m$^{-2}$. Note that the RM
structure in the diffuse radiation is of the same order of magnitude
but uncorrelated to the RM's of the extragalactic sources.

\section{Interpretation of the observations}
\label{sec_model}


The galactic synchrotron radiation is emitted in a halo of
relativistic electrons with a scale height $\sim$3.6 kpc at the radius
of the Sun (the ``thick disk'' in \cite{beuermann}). The synchrotron
emissivity is strongest in the galactic plane, and decreases
outwards.

Reynolds determined the exponential scale height of thermal electrons
in the galaxy to be $h_e \approx 1.5$ kpc, forming the so-called
Reynolds layer \cite{reynolds}. Thus at several kpc above the galactic
plane, the thermal electron density is very low and Faraday rotation
is negligible, whereas in the Reynolds layer considerable Faraday
rotation occurs.

The galactic magnetic field consists of a large-scale component, which
at this longitude ($l = 161^{\circ}$) is almost perpendicular to the
line of sight, and of a random component \cite{han}. Therefore,
$B_{\perp}$ (determining the intensity of the synchrotron emission)
includes the complete large-scale galactic magnetic field, whereas
$B_{\parallel}$ (determining Faraday rotation) is almost completely
random.

The non-detection of structure in synchrotron emissivity puts
constraints on the structure in relativistic electron density
$n_{e,rel}$ and magnetic field perpendicular to the line of sight,
$B_{\perp}$. Absence of structure in $n_{e,rel}$ and magnetic field in
the halo is also observed in external galaxies, e.g. M51 \cite{berk}
and M31 \cite{han2}.  For total intensity $I$ to be constant in the
Reynolds layer as well, $n_{e,rel}$ must be smooth, and the
large-scale magnetic field component must dominate the turbulent
component.  If the Reynolds layer consists of many cells with
different turbulent magnetic field $B_{\perp,turb}$, some local
variations in synchrotron emission could exist, which would be
averaged out along the line of sight. However, a large amount of
turbulent cells also decreases the degree of polarization. As we
observe a high degree of polarization, the amount of possible
turbulent cells is limited, and only a few K in local variations in
total intensity could be erased in this way.

So the polarized component of the synchrotron emission also has little
or no small-scale structure. However, Faraday rotation in the Reynolds
layer varies on small scales (Fig.~\ref{fig2}), inducing structure
in polarization angle. Linearly polarized radiation emitted at
different depths is Faraday-rotated by different amounts, which causes
depolarization (internal Faraday dispersion \cite{gardner}) on small
scales, and so induces small-scale structure in $P$. Emission from the
far side of the Reynolds layer is mostly depolarized, so the polarized
emission we observe comes from the nearest part of the layer.

Therefore, the observed Rotation Measure is built up in this nearest
part of the layer, and structure in RM denotes structure in
$B_{\parallel}$ in the nearest part of the layer. The observed RM in
extragalactic sources, however, is built up over the entire line of
sight. The gradient in RM of the sources on degree scales in
Fig.~\ref{fig1}b, roughly in the direction of galactic latitude, is
not visible in the RM structure of the diffuse radiation, i.e. not
dominant in the nearest part of the Reynolds layer.

Concluding, our observations of total intensity $I$, polarized
intensity $P$, and Rotation Measure of diffuse radiation and
extragalactic sources, impose many constraints on the possible nature
of magnetic field and electron density in the warm ionized gas and in
the synchrotron halo (for details and calculations, see
\cite{haverkorn2}).

\acknowledgements{The Westerbork Synthesis Radio Telescope is operated
by the Netherlands Foundation for Research in Astronomy (ASTRON) with
financial support from the Netherlands Organization for Scientific
Research (NWO). This work is supported by NWO grant 614-21-006.}

\vfill

\begin{iapbib}{99}{
\bibitem{berk} Berkhuijsen E. M., Horellou, C., Krause, M. \et, 1997,
        \aeta 318, 700
\bibitem{beuermann} Beuermann K., Kanbach G., Berkhuijsen E. M., 1985,
	\aeta 153, 17
\bibitem{han} Han J. L., 2000, eds Strom R. \& Nan R.-D.,
 	in {\it proceedings of IAU Coll. 182}, in prep.
\bibitem{han2} Han J. L., Beck R., Berkhuijsen E. M., 1998, \aeta 335, 1117
\bibitem{gardner} Gardner F. F., Whiteoak J. B., 1966, ARAA 4, 245 
\bibitem{haverkorn} Haverkorn M., Katgert P., de Bruyn, A. G., 2000,
	\aeta 356, L13
\bibitem{haverkorn2} Haverkorn M., Katgert P., de Bruyn, A. G.,
	submitted to \aeta
\bibitem{leahy} Leahy J. P., 1987, \mn 226, 433
\bibitem{reynolds} Reynolds R. J., 1989, \apj 339, 29L
\bibitem{wieringa} Wieringa M. H., de Bruyn A. G., Jansen D. \et,
	1993, \aeta 268, 215 }
\end{iapbib}
\end{document}